# THE DARHT PHASE 2 LINAC*

H. L. Rutkowski, L. L. Reginato, W.L. Waldron, K. P. Chow, M. C. Vella, W. M. Fawley,
Lawrence Berkeley National Laboratory, 1 Cyclotron Road, Berkeley, CA 94720
R. Briggs, Science Applications International Corp., 7041 Koll Center Parkway, Suite 260
Pleasanton, CA 94566
S. Nelson, Lawrence Livermore National Laboratory, 7000 East Avenue, Livermore, CA 94550
Z. Wolf, Stanford Linear Accelerator Center, 2575 Sand Hill Rd., Menlo Park, CA 94025
D. Birx, Science Research Laboratory, 15 Ward Street, Somerville, MA 02143

*Abstract*

The second phase accelerator for the Dual Axis Hydrodynamic Test facility (DARHT) is designed to provide an electron beam pulse that is 2μs long, 2kA, and 20 MeV in particle energy. The injector provides 3.2 MeV so that the linac need only provide 16.8 MeV. The linac is made with two types of induction accelerator cells. The first block of 8 cells have a 14 in. beam pipe compared to 10 in. in the remaining 80 cells. The other principal difference is that the first 8 cells have reduced volt-sec in their induction cores as a result of a larger diameter beam pipe. The cells are designed for very reliable high voltage operation. The insulator is Mycalex. Results from prototype tests are given including results from solenoid measurements. Each cell contains a solenoid for beam transport and a set of x-y correction coils to reduce corkscrew motion. Details of tests to determine RF mode impedances relevant to BBU generation are given. Blocks of cells are separated by "intercells" some of which contain transport solenoids. The intercells provide vacuum pumping stations as well. Issues of alignment and installation are discussed.

## 1 INTRODUCTION

The Dual-Axis Radiographic Hydrodynamic Test (DARHT) facility at Los Alamos National Laboratory (LANL) is a pulsed X-ray radiography machine for the national Stockpile Stewardship Program. It consists of two linear accelerators oriented at 90° to each other and with both pointed at the same target. The first linac (Phase 1) is a short pulse single shot accelerator using ferrite induction cells. The Phase 2 linac is a long pulse linac using Metglas induction cells. The beam at the exit of the accelerator is required to be 2 kA of electrons at 20 MeV and with a 2 μs long flat current pulse. Downstream kicker systems being designed and built by Lawrence Livermore National Laboratory (LLNL) will select 4 short pulses out of the 2 μs macropulse for delivery to the X-ray conversion target. The physics design issues that dominate the design are transverse RF mode impedances and Q's that can generate BBU, transport solenoid field alignment and energy flatness both of which contribute to corkscrew motion of the beam centroid, vacuum quality, and emittance growth. The engineering issues that dominate the design are packing enough volt-seconds of induction core material into a cell within the pre-set building size constraints, high voltage reliability of the systems, mechanical stability and alignment, and providing sufficient space for diagnostics, sufficient maintenance capability, and adequate vacuum pumping. The injector provides a 3.2 MeV, 2kA beam to the accelerator in a pulse with 2.1 μs flat current and a 500 ns rise time (0-99%). The linac consists of 88 accelerator cells adding 16.8 MeV to the beam energy. The first 8 cells are called "injector cells" and are designed to transport the entire injector pulse to a Beam-head Clean Up Zone (BCUZ), which chops off the pulse rise time. This beam element is designed and built by LANL. The rest of the linac consists of 80 "standard cells", which are in blocks of 6 except for a final block of 8. The cellblocks are separated by removable elements called intercells. The beam exiting the linac should have transverse motion of the centroid from all causes less than 10% of the beam radius (5 mm) and normalized emittance of no more than 1000 π mm-mrad.

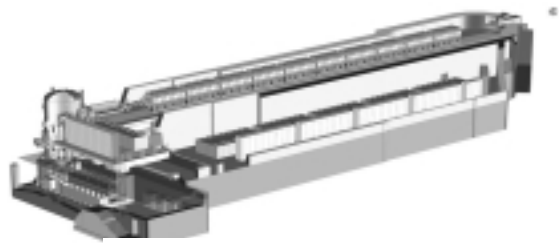

Figure 1 - DARHT Phase 2 Linac

## 2 ACCELERATOR INDUCTION CELLS

A drawing of the standard accelerator cell is shown in fig. 2. The beam tube has an inner diameter of 10 inches as chosen from preliminary BBU and beam loss considerations. The induction cores are Honeywell (Allied Signal) 2605-SC Metglas with a specified v-s product of 0.48 v-s per cell. Actual delivered cores are providing up to 0.51 v-s. Each cell contains 4 pancake sub-cores and the entire ensemble is driven as a single unit by a single pulser. The cores are assembled into the aluminium cell housing and then vacuum pumped before the cell is filled with Shell Diala high voltage oil. The beam tube is 304 stainless steel and is hollow to accommodate a transport solenoid immersed in a water-cooling jacket.

*Work supported by the US Department of Energy under contract DE-AC03-76SF00098

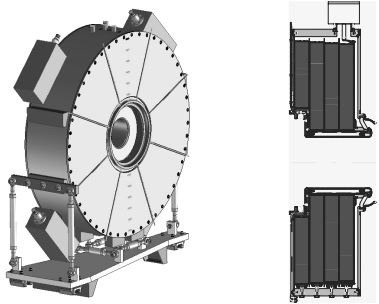

Figure 2 - Standard Accelerator Cell

The solenoid is a wound copper coil with polyester-amide-imide insulated conductor wet-wound and painted in Castall 301 epoxy. Each coil has 2400 turns in 12 layers and can be driven to provide 0.2T field on the beam tube axis within the limits of the cooling system provided by LANL. The solenoids will be operated in a pulsed mode except for the injector cell solenoids.

The insulator is Mycalex (dielectric constant 6.7) and is a structural member of the cell. It is a mica-glass composite chosen for its high strength without brittleness and its excellent vacuum and high voltage properties. Its high dielectric constant was initially thought to present TM mode impedance problems and considerable design effort using the AMOS code was necessary to find an optimal structure and ferrite damping design. Outside the beam tube and in the oil-filled section, flexible PC board type corrector coils, both vertical and horizontal, are wrapped around the tube. The entire cell is supported by a 6 strut mounting system developed at Lawrence Berkeley National Laboratory (LBNL) for the Advanced Light Source. It uses differential screws to allow precision movement of the heavy cell (7 tons) during alignment. The cells are vacuum-sealed to each other using a clamp ring-bellows system that allows movement for alignment. Each accelerator cell will be fiducialized with respect to the magnetic axis of its solenoid using a standard stretched wire technique by LANL. In that way, each cell can be aligned with respect to the ideal beam axis as it is installed in the linac. The magnetic axis of each solenoid is to be within 400 μ of the ideal beam axis (3σ) for offset and within 3 mrad (3σ) for tilt in both Euler angles. The injector cells are a scaled version of the standard cells. The beam tube ID was enlarged to 14 in. while keeping the outer cell dimensions constant. The main effects are to enlarge the beam tube, solenoid, and insulator while sacrificing Metglas volume on the inside of the cores. Consequently, the injector cells operate at 175 kV pulse voltage compared to 193 kV for the standard cells to maintain constant pulse length. The beam tube enlargement allows full transmission of the injected pulse to the BCUZ and takes advantage of the well known scaling ($1/b^2$ where b is the tube radius) for BBU in pillbox cavities[1]. Three prototypes were constructed as part of the design. The first, without a solenoid, was built in two forms for high voltage testing and RF transverse mode measurements at LBNL. The second was an earlier version of the injector cell design while the third was the final standard cell design. Both were sent to the THOR facility at LANL. The cells have shown excellent high voltage performance and each standard production cell is tested at 200 kV for 2000 shots before shipping.

## 3 TRANSVERSE MODE MEASUREMENTS

Measurements using both the standard TSD[2] method for highly damped cavities and a new two-wire excitation - loop pickup method invented by two of the authors (Briggs and Birx) were used. The optimal damping geometry and ferrite placement was chosen by a computer design activity using AMOS. The standard cell measurements were carried out initially on the first prototype cell and finalized on the first production standard cell. The dipole transverse mode frequencies were found to be 170, 230, and 577 MHz with real impedances of 182, 259, and 283 ohms per m respectively. These values were found to be sufficiently low by the integrated beam dynamics team. The two methods were found to agree within experimental error so that the two wire-loop technique was used by itself on the injector cell. The injector cell frequencies were 152 and 200 MHz shifted from the standard cell and the impedances were 152 and 149 ohms per m, well within the limits established by the beam dynamics studies. The damping ferrite used was CMI Technology, tile material, N2300, distributed azimuthally around the beam tube on the oil side in pie sections, which were separated to avoid saturation by the driver pulse current.

## 4 MAGNETIC CHARACTERIZATION MEASUREMENTS

The solenoids were designed to be fine wire magnets using many turns. This design eliminates the need to package an internally cooled bulky conductor in a very confined space, reduces power consumption, and provides very high quality field. Several (10) solenoids have been characterised at Stanford Linear Accelerator Center (SLAC) prior to their installation in accelerator cells. The solenoids are first installed in their beam tube housings, which are welded shut. They are then sent to SLAC where they are aligned using standard SLAC measurement procedures and their fields are measured. A rotating coil probe method is used, which gives data on the transverse field components as a function of distance along the axis in the solenoid. This data is then processed in a fitting routine at LBNL to find the effective magnetic axis of the solenoid. A few solenoids have also been checked for sextupole fields, which have turned out to be negligible. If a large effective tilt (>1.0 mrad) is detected in one run, the solenoid is realigned and rechecked. The largest effective tilt between the magnetic and mechanical solenoid axes in the set of measurements taken with both standard cell solenoids and injector cell solenoids is 0.70 mrad. The range is 0.03-0.7 mrad. After field measurements, the solenoids are returned to LBNL for assembly into accelerator cells.

## 5 INTERCELLS

The cellblocks in the linac are separated by intercells, which provide a removable element that allows sliding the interlocking accelerator cells apart for removal in case of maintenance. They also allow for diagnostic ring removal at each cellblock, insertion of intercepting diagnostics, and vacuum pumping. A cross-section drawing of an azimuthally symmetric intercell is shown in fig. 3. The diagnostic ring, designed by LANL, is shown between the downstream bellows and the housing of the downstream cell. The intercell body provides the positive side of the accelerating gap for the upstream cell whose beam pipe and insulator are shown. This figure shows an intercell with a transport solenoid. Only the first 4 intercells have these solenoids, which are needed for beam transport matching to reduce emittance growth and BBU. In the other 5 intercells, the solenoid is absent and the extra axial space is used for additional pumping speed. Twelve current return bars complete the circuit across the pumping throat opening to smooth the B field of the return current. Each opening is filled with 95% open area stainless steel mesh to shield the intercell cavity from beam generated RF. The intercells are mounted on the upstream cell with turnbuckle struts and the weight is supported by a single point suspension "pogo stick". In this way, the intercell can be independently aligned to the ideal abeam line. Vacuum modelling, backed up by empirically determined outgassing rates on actual production cells, yields an average background pressure on axis of $< 1.5 \times 10^{-7}$ Torr between the BCUZ exit and the exit of the first four cell blocks of standard cells. It is approximately $0.8 \times 10^{-7}$ Torr after that due to the higher pumping speed available after the intercell solenoids are eliminated. This should be adequate to eliminate ion hose instability.

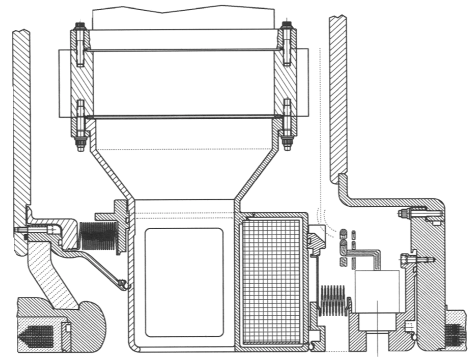

Figure 3 - Intercell

## 6 CONCLUSION

A high quality design for DARHT Phase 2 linac accelerator cells has been achieved and cells are in production. These cells have been shown to have excellent high voltage performance even with some beam loss, and measurements have shown the RF mode impedances relevant to BBU are sufficiently low to ensure meeting design requirements. The background pressure has been reduced by good vacuum system design to safe limits. Also, a very high quality solenoid field has been achieved within operational and cooling system constraints. Finally, a design has been achieved which delivers the energy and pulse length required in spite of serious space restrictions.